# Measurement of magnetic field induced by spin- accumulated electrons in a FeCoB nanomagnet

Vadym Zayets

National Institute of Advanced Industrial Science and Technology, Tsukuba, Ibaraki 305, Japan

**The spins, which are accumulated at a boundary of a nanomagnet due to the Spin Hall effect, induce a magnetic field, which tilts the nanomagnet magnetization out of its easy axis. Even though this magnetic field is relatively small (about 20 Gauss), it can reverse the magnetization of the nanomagnet if it is modulated in resonance with the magnetization precession and the conditions of the parametric resonance are met. Therefore, such magnetic field can be used to optimize the recording mechanism and to minimize the recording energy for MRAM. A high-precession measurement method of the spin-accumulation-induced magnetic field is developed. The magnetic field is measured in a FeCoB nanomagnet and its properties are studied.**

## I. INTRODUCTION

The reduction of energy consumption is one of the major challenges of present MRAM development. The recording mechanisms of the Spin Torque (ST) MRAM and the Spin-Orbit Torque (SOT) MRAM are based on the spin injection from the "pin" to "free" layer of an MRAM cell. A possible reduction of the MRAM recording power is limited by a minimum amount of the injected spin-polarized electrons, which is necessary to create a sufficient ST or SOT torque in order to reverse the magnetization of the "free" layer [1]–[6]. This fundamental limitation is difficult to avoid. The reduction of the volume of the "free" layer is only a feasible method for the reduction of the MRAM recorded power.

Recently, a new recording mechanism was proposed for a MRAM cell, which is based on a parametric resonance [7]. This mechanism requires a substantially smaller spin torque for the magnetization reversal due to its resonance nature and, therefore, it can be a solution for the further reduction of the recording energy for MRAM operation.

There are several possible mechanisms [7] for the parametric magnetization reversal (PMR). A commonly required feature for such a mechanism is the ability to modulate the magnetization direction of a nanomagnet by an electrical current flowing through the nanomagnet. One possible PMR mechanism is due to the current-induced magnetic field $H^{(CI)}$, the direction of which is perpendicular to the nanomagnet magnetization [7]. The magnetic field with the required properties can be created by accumulated spin-polarized electrons (see Fig.1). The spin accumulation is created by an electrical current at a nanomagnet boundary due the Spin Hall effect [8]. As a result, the amount of spin accumulation and, therefore, the magnitude of $H^{(CI)}$ are proportional to the electrical current, as is required for the PMR.

One origin of the Spin Hall effect is a spin-dependent scattering of a conduction electron, which symmetry is related to the crystal property of the ferromagnetic and non-magnetic metals and the properties of the interface between them. As a result, the spin direction of the spin accumulation, which is created due to the Spin-Hall-effect, may be in both directions:

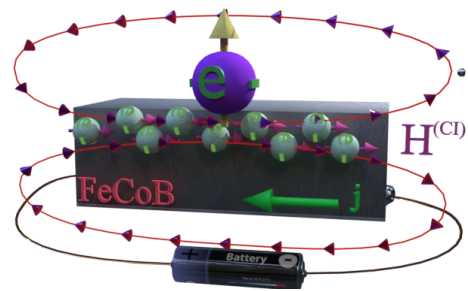

Fig. 1. An electrical current, which flows through FeB nanomagnet, creates the spin accumulation (green balls). The magnetic field $H^{(CI)}$, which is created by the spin accumulation, forces the magnetization (blue ball) to tilt from equilibrium perpendicular direction to in-plane direction.

perpendicular-to-plane and in-plane. Each type of spin accumulation can be used for the PMR. This Letter only studies only the in-plane component magnetic field $H^{(CI)}$ induced by the spin accumulation. The tilting of the magnetization from the easy axis due to this field component is the mechanism used for the PMR. The perpendicular-to-plane field component can also induce the PMR, but the mechanism involved is different and is not described in this Letter. It should be noted that the $H^{(CI)}$ is the effective magnetic field, which includes the effect of the exchange interaction between spin-polarized conduction electrons and the localized d-electrons.

This Letter describes a new high-precision measurement method for the magnetic field $H^{(CI)}$ induced by spin accumulation in a nanomagnet. The dependencies of $H^{(CI)}$ on the electrical current and an external perpendicular magnetic field $H_z$ are studied in a $Fe_{0.4}Co_{0.4}B_{0.2}$ nanomagnet with perpendicular-to-plane equilibrium magnetization. The $Fe_{0.4}Co_{0.4}B_{0.2}$ nanomagnet[10]–[14] is the most- used ferromagnetic material in the present MRAM .

## II. MEASUREMENT METHODS OF CURRENT INDUCED MAGNETIC FIELD $H^{(CI)}$

### A. Measurement scheme and challenges

A measurement of a magnetic field, which is induced by a tiny nanomagnet, is a challenging task. An additional

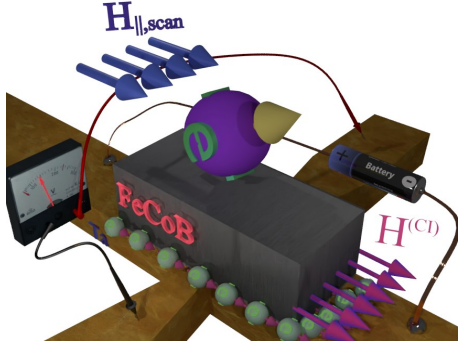

Fig. 2. Proposed measurement method. The Hall voltage $V_{Hall}$ is measured under a DC current. Under a scanned external magnetic field $H_{\parallel,scan}$ and spin-accumulation- induced magnetic field $H^{(CI)}$, nanomagnet magnetization (blue ball) is tilted towards the in-plane direction and the Hall voltage $V_{Hall}$ is reduced. The $H^{(CI)}$ is evaluated from assymetry of measured $V_{Hall}$ with respect to polarity reversal of $H_{\parallel,scan}$

measurement challenge is that the measured magnetic field $H^{(CI)}$ is very small. Typically, it is only about a few Gauss. (See Fig.4).

When a magnetic field is applied perpendicularly to the easy axis of a nanomagnet, the nanomagnet magnetization is tilted towards the applied field. The tilting angle and consequently the $H^{(CI)}$ are measured by a Hall setup. The Hall angle of the Anomalous Hall effect (AHE) is linearly proportional to the perpendicular component $M_z$ of the magnetization M. When the magnetization is tilted, the $M_z$ is calculated as [15]

$$\frac{M_z}{M} = \sqrt{1 - \left(\frac{H^{(CI)}}{H_{ani}}\right)^2} \qquad (1)$$

where $H_{ani}$ is the anisotropy field. The measured $H_{ani}$ in our FeCoB nanomagnet is ~ 8 kGauss and the measured $H^{(CI)}$ is ~20 Gauss (See Fig.4). This means that the tilted angle is only 140 mdeg and the change of the Hall angle and therefore the Hall voltage is only ~ 0.0003 % (Eq.(1)). It is a challenge to measure such a small change of the Hall voltage in a nanomagnet.

### B. Measurement method of the 2nd harmonic

The measurement method of the 2nd harmonic [16]–[18] is used to estimate the spin torque of the STT and/or the SOT type and to measure the angle of the magnetization precession induced by the spin torque torque. The method also measures $H^{(CI)}$.

The method of the 2nd harmonic measures the amount of modulation of the perpendicular component $M_z$ of the magnetization M by the electrical current. In the method the electrical current in the nanomagnet is modulated at a low frequency (~ 1 kHz) and the second harmonic of the Hall voltage is measured by the lock-in technique.

The Hall voltage is linearly proportional to the current j and to the perpendicular component of the magnetization $M_z$ [19]. When the $M_z$ is affected by j, both the $M_z$ and j become modulated and the frequency beating between them creates the second harmonic of the Hall voltage, from which the dependency of $M_z$ on j is evaluated.

There are only three possible effects, which may cause the modulation of $M_z$ by the electrical current j. The first effect is the heating of the nanomagnet by the current. When the nanomagnet temperature increases the nanomagnet magnetization decreases according to the Curie-Weiss law. Even though low frequency modulation is used in this method, it is sufficient to assume that there is no temperature modulation by the current and this contribution can be ignored.

The second effect is the magnetization precession. When the electrical current induces the spin torque, there is a magnetization precession round the easy axis (the z-axis) and the $M_z$ is reduced as:

$$\frac{M_z}{M} = \cos(\theta) \qquad (2)$$

where θ is the precession angle, which depends on the spin torque and therefore on the current j. The measurement of θ and the evaluation of the spin torque is the main desired goal of the measurement method of the 2nd harmonic[17], [18].

The third effect, which contributes to the current modulation of $M_z$, is the magnetization tilting due to $H^{(CI)}$ as described by Eq.(1) and shown in Fig.2. In the method of 2nd harmonic, this contribution is undesirable and makes the interpretation of the measured data ambiguous.

In the proposed method, which is described below, $H^{(CI)}$ is measured without any influence of the magnetization precession and the spin torque. As a result, in the proposed measurement method there is no ambiguity between contributions of the spin torque and $H^{(CI)}$. The combination of both measurement methods might resolve the problem of a systematic error due to $H^{(CI)}$ in the measurement of the spin torque in the measurement method of the 2nd harmonic.

### C. Proposed measurement method

In the proposed method, the Hall voltage is measured at a DC current by a nanovoltmeter (Fig.2) under a scanned in-plane external magnetic field $H_{\parallel,scan}$. Since the easy axis of the studied nanomagnets is perpendicular-to-plane, an in-plane magnetic field tilts the magnetization towards the in-plane

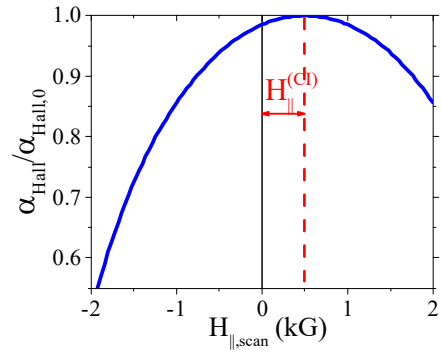

Fig. 3. Measurement principle. The Hall angle vs. scanned in-plane magnetic field $H_{\parallel,scan}$. The dependence is asymmetric versus a reversal of $H_{\parallel,scan}$ due to non-scanned magnetic field $H^{(CI)}$. The $H^{(CI)}$ is evaluated by minimizing difference between positive and negative parts of the data with an offset.

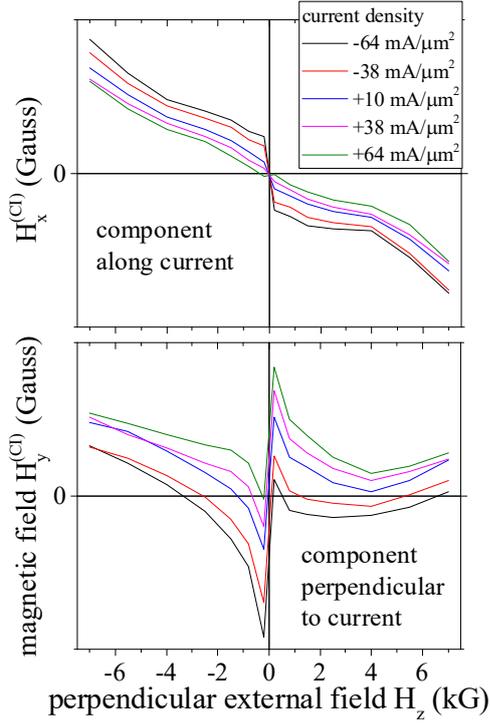

Fig.4 Measured spin-accumulation-induced magnetic field $H^{(CI)}$ as a function of a perpendicular-to- plane magnetic field $H_z$ and current density. Two in-plane components of $H^{(CI)}$ are shown. Lines of different color corresponds to a different current density.

direction and the Hall angle $\alpha_{Hall}$ is reduced according to Eq.(1).

In the absence of $H^{(CI)}$, the dependence of the Hall angle $\alpha_{Hakl}$ on $H_{\parallel,scan}$ is independent of the polarity of $H_{\parallel,scan}$:

$$\alpha_{Hall}\left(H_{\parallel,scan}\right) = \alpha_{Hall}\left(-H_{\parallel,scan}\right) \qquad (3)$$

because of the symmetry of the measurement.

When there is an additional in-plane magnetic field, for example the $H^{(CI)}$, which is not scanned, the symmetry of Eq.(3) becomes (See Fig.3)

$$\alpha_{Hall}\left(H_{\parallel,scan} - H^{(CI)}\right) = \alpha_{Hall}\left(-H_{\parallel,scan} + H^{(CI)}\right) \quad (4)$$

From Eq.(4) the in-plane magnetic field $H^{(CI)}$ can be evaluated using the measured dependence $\alpha_{Hall}(H_{\parallel,scan})$ (See Fig.3) by minimizing the expression:

$$\left[\alpha_{Hall}\left(H_{\parallel,scan} - H^{(CI)}\right) - \alpha_{Hall}\left(-H_{\parallel,scan} + H^{(CI)}\right)\right]^2 \quad (5)$$

The estimated measurement precision of $H^{(CI)}$ in a measured FeCoB nanomagnet is about 0.1- 1 Gauss depending on the sample.

The proposed method is a direct measurement method of $H^{(CI)}$. As a result, any possible systematic error, if it exists, can be easily identified. It should be noted that the measurement of $H^{(CI)}$ does not depends on the explicit form of the dependence of $\alpha_{Hall}$ on $H_{\parallel,scan}$ (Eq.1). This makes the measurement method even more robust against a possible systematic error.

## III. MEASUREMENT OF $H^{(CI)}$ IN A FeCoB NANOMAGNET

The magnetic field $H^{(CI)}$ is measured in a 1.1-nm-thick $Fe_{0.4}Co_{0.4}B_{0.2}$ fabricated on top of a 2.5-nm-thick Ta nanowire contacted by a pair of Hall probes. The equilibrium magnetization of the nanomagnet is perpendicular-to-plane. The details of the sample geometry and sample preparation are described in [9]. The magnetic and structural properties of $Fe_{0.4}Co_{0.4}B_{0.2}$ are described in [10]–[14]. About 100 nanomagnets, the sizes of which vary between 50 x 50 nm and 3000 x 3000 nm, were measured. The measured data for nanomagnets of different sizes are systematic. The external in-plane magnetic field is scanned either along or perpendicular to the current for an individual measurement of each of the two in-plane components of $H^{(CI)}$. Additionally, a perpendicular-to-plane external magnetic field $H_z$ is applied in order to clarify the origin and properties of $H^{(CI)}$. For each value of $H_z$, the in-plane magnetic field $H_{\parallel,scan}$ is independently scanned and the value of $H^{(CI)}$ is evaluated. At each value of $H_z$, and $H_{\parallel,scan}$, the nanomagnet was in a single-domain state, which was confirmed by independent measurements [9], [20].

Figures 4-6 show the measured $H^{(CI)}$ as a function of $H_z$ and the current density j and its current derivative for a 3000 x 3000 nm nanomagnet. The $H^{(CI)}$ depends substantially on both the $H_z$ and the current density j and both dependencies are related to each other.

## IV. DISCUSSION

It was suggested [17] that the method of the $2^{nd}$ harmonic measures two different types of torques. The field-like (FL) torque is measured when the magnetization is inclined along the electrical current. The damp-like (DL) torque is measured when the magnetization is inclined perpendicularly to the current. The suggestion was based on the measured symmetry of the measured $2^{nd}$ harmonic with respect to the magnetization reversal. When the magnetization is inclined along the current under an external magnetic field, the measured $2^{nd}$ harmonic reverses its polarity when $M_z$ is reversed. Therefore, it is assumed that the measured spin

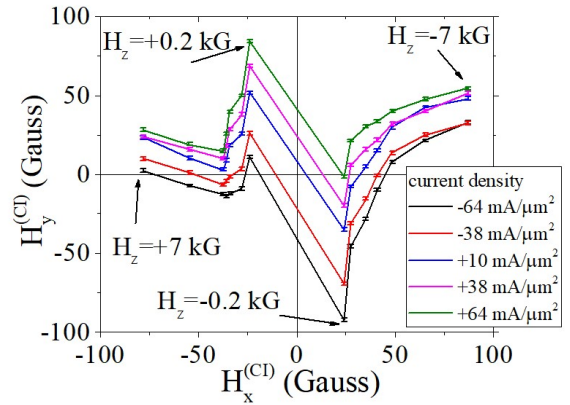

Fig.5 Component $H_x^{(CI)}$ along current vs. component $H_y^{(CI)}$ perpendicular to current. The magnetic field $H_z$ is used a parameter and scanned from -7 kGauss to +7 kGauss. The measured data are the same the data of Fig. 4.

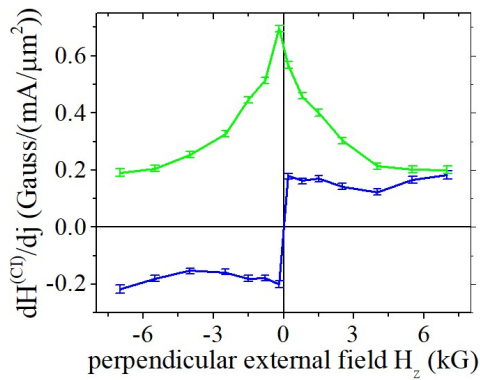

Fig.6 Components of derivative $dH^{(Cl)}/dj$ along and perpendicular to current $j$.

torque is proportional to $M^1$ and, therefore, corresponds to the field-like torque:

$$\left(\frac{\partial \vec{M}}{\partial t}\right)_{FL} \sim \vec{M} \times \vec{j} \qquad (6)$$

The symmetry of the measured $2^{nd}$ harmonic is different when the magnetization is inclined perpendicularly to the current under an external perpendicular-to-current in-plane magnetic field. In this case the $2^{nd}$ harmonic keeps its polarity when $M_z$ is reversed. Therefore, it is assumed that the measured spin torque is proportional to $M^2$ and, therefore, corresponds to the damp-like torque:.

$$\left(\frac{\partial \vec{M}}{\partial t}\right)_{DL} \sim \vec{M} \times \left(\vec{M} \times \vec{j}\right) \qquad (7)$$

We have observed the same symmetry for the measured $H^{(Cl)}$ (Figs.4 and 6). When $M_z$ is reversed, the component of $H^{(Cl)}$ along the current reverses its polarity (Fig.4a), but the component perpendicular to the current keeps its polarity (Fig.4b). Such symmetry is not related to the symmetries of the FL or DL torques.

It is clear from Fig.5, that the nanomagnet experiences two types of the in-plane magnetic field; the current-dependent $H^{(Cl)}$ field and the current-independent $H^{(0)}$ field. The direction of $H^{(0)}$ is fixed to the direction of $H^{(0)}$. When the magnetization is reversed at the coercive field $H_c$, the field $H^{(0)}$ is also switched. That is why there are two mirror-like parts in Fig.5.

Additionally, the dependencies in Fig.5 indicate that the field $H^{(0)}$ has a substantial perpendicular-to-plane component. Even though this component is not measured, it clearly reverses its direction when the magnetization is reversed. As a result, the measured in-plane component of $H^{(0)}$ rotates 180 deg. The current-dependent magnetic field $H^{(Cl)}$ fully follows the trajectory of the $H^{(0)}$ rotations. The physical origin of such complex behavior yet has to be understood.

## V. Conclusion

When an electrical current flows through the nanomagnet, the spins are accumulated at the nanomagnet boundaries due to the Spin Hall effect. There is a magnetic field, which is induced by this spin accumulation. The field is small but measurable. The importance of this field is that if it is modulated in resonance with the magnetization precession, it can reverse the nanomagnet magnetization. This parametric mechanism is beneficial for a low-energy MRAM recording.

A high-precision measurement method for the magnetic field, which is induced by the accumulated spins, is proposed and developed. The magnetic field is measured in a FeCoB nanomagnet and its properties are studied.

It is noted that the commonly-used measurement of the $2^{nd}$ harmonic may have a substantial systematic error due to the studied magnetic field $H^{(Cl)}$. It is shown that there is a substantial contribution of $H^{(Cl)}$ to the $2^{nd}$ harmonic of the Hall voltage and this contribution has the same symmetry as the symmetry used to distinguish between the field-like torque and the damp-like torque. The contribution of $H^{(Cl)}$ should be considered when the spin torque is evaluated from the $2^{nd}$ harmonic measurement.

## References

[1] M. Cubukcu *et al.*, "Ultra-Fast Perpendicular Spin-Orbit Torque MRAM," *IEEE Trans. Magn.*, vol. 54, no. 4, pp. 1–4, 2018, doi: 10.1109/TMAG.2017.2772185.

[2] T. Valet and A. Fert, "Theory of the perpendicular magnetoresistance in magnetic multilayers," *Phys. Rev. B*, vol. 48, no. 10, pp. 7099–7113, Sep. 1993, doi: 10.1103/PhysRevB.48.7099.

[3] R. Fiederling *et al.*, "Injection and detection of a spin-polarized current in a light-emitting diode," *Nature*, vol. 402, no. 6763, pp. 787–790, Dec. 1999, doi: 10.1038/45502.

[4] S. A. Crooker and D. L. Smith, "Imaging spin flows in semiconductors subject to electric, magnetic, and strain fields," *Phys. Rev. Lett.*, vol. 94, no. 23, p. 236601, Jun. 2005, doi: 10.1103/PhysRevLett.94.236601.

[5] S. A. Crooker *et al.*, "Applied physics: Imaging spin transport in lateral ferromagnet/ semiconductor structures," *Science (80-. ).*, vol. 309, no. 5744, pp. 2191–2195, Sep. 2005, doi: 10.1126/science.1116865.

[6] Y. Ohno, D. K. Young, B. Beschoten, F. Matsukura, H. Ohno, and D. D. Awschalom, "Electrical spin injection in a ferromagnetic semiconductor heterostructure," *Nature*, vol. 402, no. 6763, pp. 790–792, Dec. 1999, doi: 10.1038/45509.

[7] V. Zayets, "Mechanism of parametric pumping of magnetization precession in a nanomagnet. Parametric mechanism of current-induced magnetization reversal," Apr. 2021, Accessed: Aug. 03, 2021. [Online]. Available: https://arxiv.org/abs/2104.13008v1.

[8] Y. K. Kato, R. C. Myers, A. C. Gossard, and D. D. Awschalom, "Observation of the Spin Hall Effect in Semiconductors," *Science (80-. ).*, vol. 306, no. 5703, pp. 1910–1913, Dec. 2004, doi: 10.1126/science.1094383.

[9] V. Zayets and A. S. Mishchenko, "Hall effect in ferromagnetic nanomagnets: Magnetic field dependence as evidence of inverse spin Hall effect contribution," *Phys. Rev. B*, vol. 102, no. 10, p. 100404, Sep. 2020, doi: 10.1103/PhysRevB.102.100404.

[10] D. García, J. L. Muñoz, G. Kurlyandskaya, M. Vázquez, M. Ali, and M. R. J. Gibbs, "Magnetic domains and transverse induced anisotropy in magnetically soft CoFeB amorphous thin films," *IEEE Trans. Magn.*, vol. 34, no. 4 PART 1, pp. 1153–1155, 1998, doi: 10.1109/20.706424.

[11] F. Xu, Q. Huang, Z. Liao, S. Li, and C. K. Ong, "Tuning of magnetization dynamics in sputtered CoFeB thin film by gas pressure," *J. Appl. Phys.*, vol. 111, no. 7, p. 07A304, Feb. 2012, doi: 10.1063/1.3670605.

[12] J. M. D. Coey and D. H. Ryan, "Current trends in amorphous magnetism," *IEEE Trans. Magn.*, vol. 20, no. 5, pp. 1278–1283, 1984, doi: 10.1109/TMAG.1984.1063532.

[13] V. Sokalski, M. T. Moneck, E. Yang, and J. G. Zhu, "Optimization of Ta thickness for perpendicular magnetic tunnel junction

applications in the MgO-FeCoB-Ta system," *Appl. Phys. Lett.*, vol. 101, no. 7, Aug. 2012, doi: 10.1063/1.4746426.

[14] M. Yamanouchi *et al.*, "Dependence of magnetic anisotropy on MgO thickness and buffer layer in Co20Fe60B20-MgO structure," in *Journal of Applied Physics*, Apr. 2011, vol. 109, no. 7, p. 07C712, doi: 10.1063/1.3554204.

[15] M. T. Johnson, P. J. H. Bloemen, F. J. A. Den Broeder, and J. J. De Vries, "Magnetic anisotropy in metallic multilayers," *Reports Prog. Phys.*, vol. 59, no. 11, pp. 1409–1458, 1996, doi: 10.1088/0034-4885/59/11/002.

[16] U. H. Pi *et al.*, "Tilting of the spin orientation induced by Rashba effect in ferromagnetic metal layer," *Appl. Phys. Lett.*, vol. 97, p. 162507, 2010, doi: 10.1063/1.3502596.

[17] K. Garello *et al.*, "Symmetry and magnitude of spin-orbit torques in ferromagnetic heterostructures," *Nat. Nanotechnol.*, vol. 8, pp. 587–593, 2013, doi: 10.1038/nnano.2013.145.

[18] J. Kim *et al.*, "Layer thickness dependence of the current-induced effective field vector in Ta|CoFeB|MgO," *Nat. Mater.*, 2013, doi: 10.1038/nmat3522.

[19] C. M. Hurd, *The Hall Effect in Metals and Alloys*. Plenum Press, 1972.

[20] V. Zayets, "Thermally activated magnetization reversal in a FeCoB nanomagnet. High-precision measurement method of coercive field, delta, retention time and size of nucleation domain," *arXiv*. 2019.